\documentclass{article}
\usepackage{spconf,amsmath,amssymb,graphicx,color,bibunits,hyperref}
\hypersetup{hidelinks}
\makeatletter
\AtBeginDocument{%
  \renewcommand\bibcite[2]{%
    \global\@namedef{b@#1}{\hyper@@link[cite]{}{cite.#1}{#2}}}}
\makeatother

\newcommand{\stateMono}[1]{\mbox{\texttt{#1}}}
\newcommand{\metricHead}[1]{{\fontsize{7}{8}\selectfont #1}}
\newcommand{\metricSub}[1]{\textsubscript{\fontsize{6}{7}\selectfont #1}}
\DeclareMathSizes{7.5}{7}{5}{5}

\title{SYNTHESIS AND EDITING OF MULTI-INSTRUMENT AUDIO MIXTURES USING SCALAR-QUANTISED LATENTS WITH MIDI SPAN CONDITIONING}
\name{Sungkyun Chang \qquad Keshav Bhandari \qquad Simon Dixon \qquad Emmanouil Benetos}
\address{Centre for Digital Music, Queen Mary University of London}

\begin{document}
\ninept
\maketitle
\begin{abstract}
Music creation often involves iterative refinement, changing selected musical details while retaining the rest. To support such refinement, we introduce SpanSynth-Edit, a flow-matching model for MIDI-guided synthesis and editing of multi-instrument audio mixtures using low-frame-rate scalar-quantised latents. MIDI Span encodes instrument-labelled note lifecycles as unordered event sets with continuous-valued attributes and pools each set into one conditioning vector per audio-latent frame. The model uses contextual audio for instrument-specific timbre guidance and supports editing by resynthesising the target region from revised MIDI. Experiments on single- and multi-instrument benchmarks show competitive performance and demonstrate within-frame onset control. We also discuss limitations of transcription-based note-adherence evaluation.
\end{abstract}

\section{Introduction}
\label{sec:introduction}
Recent advances have enabled music-audio editing from natural-language prompts~\cite{lelan2024highfidelity,zhang2025instructmusicgen}. Refining existing music, however, often requires adding, removing, or adjusting individual notes while retaining other musical content~\cite{bhandari2025improvnet}. MIDI provides explicit control over pitches, timing, velocities, and instrument assignments for such revisions. Realising these changes in audio remains challenging: instrumental sounds overlap in a mixture, and the requested note changes must be rendered without altering concurrent notes or their timbres.

To this end, we propose SpanSynth-Edit, a MIDI-guided approach to fine-grained music-audio editing. It resynthesises a selected region from revised MIDI, using contextual audio from the original recording for timbre guidance. Aligned contextual MIDI links these sounds to instruments and notes. When adding an instrument absent from the audio context, the model draws on its learned knowledge to generate its sound.

\noindent\textbf{Related work.} MIDI-to-audio models differ in whether they synthesise instrument parts independently or as a mixture. CTD~\cite{demerle2024combining}, P-MUSE~\cite{jing2026pmuse}, and TokenSynth~\cite{kim2025tokensynth} generate single-instrument tracks, while FlowSynth~\cite{yang2025flowsynth} generates individual notes for sampled instruments. SpecDiff~\cite{hawthorne2022spectrogramdiffusion}, MAC~\cite{maman2025multiaspect}, and U-MusT~\cite{jung2026umust} synthesise multi-instrument mixtures jointly, as does SpanSynth-Edit.

\mbox{SpecDiff} and U-MusT use only preceding audio for continuation, whereas CTD and TokenSynth use global timbre embeddings. MAC relies on learned, dataset-specific performer embeddings for acoustic control. P-MUSE and our model, SpanSynth-Edit, accept contextual audio from one or both sides of the target, with or without aligned contextual MIDI. Our model also renders mixtures containing instruments absent from the context.

MIDI-guided audio editing has been benchmarked for single instruments~\cite{jing2026pmuse}, while related audio-to-audio methods address timbre transfer~\cite{lee2026diffusiontimbre}. Selected MIDI-to-audio models can in principle support editing through resynthesis from revised MIDI, motivating their use as baselines. However, the challenge is to render the requested note changes or new instrument parts while preserving the notes and timbres of other concurrent parts. We also explore adapting FlowEdit~\cite{kulikov2025flowedit}, a training-free method for content-preserving image editing, to MIDI-guided audio editing.

SpanSynth-Edit combines joint multi-instrument synthesis and editing using reference context. Our contributions address three complementary aspects of this setting.

\noindent\textbf{Audio representation.} We perform flow matching with scalar-quantised (SQ) audio latents~\cite{yang2025simplespeech2,yang2026heartmula}, rather than mel-spectro\-grams~\cite{jing2026pmuse,hawthorne2022spectrogramdiffusion}, codebook tokens~\cite{kim2025tokensynth,jung2026umust,tang2025midivalle}, or unquantised latents~\cite{demerle2024combining,lee2026diffusiontimbre}. The low frame rate of SQ latents keeps sequences short to support efficient training and inference.

\noindent\textbf{Note representation.} We propose MIDI Span, a frame-aligned note representation derived from MIDI. Conventional piano rolls use one grid per instrument~\cite{jing2026pmuse,maman2025multiaspect} and quantise note timing to frames, losing precision at low frame rates. Note sequences~\cite{tang2025midivalle,borovik2025symupe} retain fine timing but serialise concurrent notes. MIDI Span instead represents each instrument-labelled note throughout its lifecycle as frame-aligned events with real-valued numerical attributes that retain within-frame timing. A permutation-invariant encoder~\cite{zaheer2017deep} pools each frame's variable-sized event set into one conditioning feature.

\noindent\textbf{Evaluation suite.} We evaluate single- and multi-instrument synthesis and editing for audio quality, similarity, and note and instrument adherence. For editing, we construct paired original and revised versions, each with ground-truth audio and MIDI, to assess adherence to requested changes and preservation of unchanged notes. The model demo, checkpoints, data, and benchmark samples are available\footnote{\label{fn:project}\href{https://mimbres.github.io/spansynth-edit/}{\textcolor{blue}{\nolinkurl{https://mimbres.github.io/spansynth-edit/}}}}.

\begin{figure*}[t]
\centering
\includegraphics[width=\textwidth]{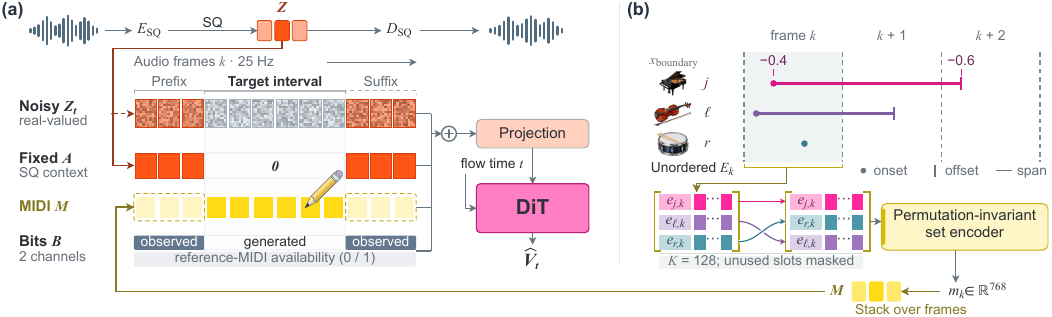}
\caption{Frame-aligned conditioning and MIDI Span. (a)~The DiT combines noisy audio with fixed audio context, MIDI features, and binary masks, with $\oplus$ denoting feature-wise concatenation. The dashed red arrow shows training interpolation, and the upper codec path shows reconstruction. (b)~Per-frame note-event sets retain within-frame timing and undergo permutation-invariant pooling.}
\label{fig:conditioning}
\end{figure*}

\section{Model}
\label{sec:model}

Given target MIDI and audio context, SpanSynth-Edit generates the audio mixture in the target region (Fig.~\ref{fig:conditioning}(a)). Observed context comprises audio and optional aligned MIDI preceding and succeeding that region. Editing resynthesises changed and unchanged notes together from revised target MIDI, without additional training.

\subsection{Scalar-Quantised Audio Representation}
\label{sec:audio_sq}

\noindent\textbf{Codec.} We use the frozen HeartCodec~\cite{yang2026heartmula} encoder $E_{\mathrm{SQ}}$ and decoder $D_{\mathrm{SQ}}$. The encoder maps 48~kHz mono audio $y$ to 128-dimensional frames at 25~Hz (hop $H=40$~ms), bounded by $\tanh$. Coordinate-wise quantisation $Q(x)=\operatorname{round}(9x)/9$ gives 19 levels in $[-1,1]$ and produces the clean SQ sequence $\mathbf Z$. The final generated latent sequence $\widehat{\mathbf Z}$ is clipped to $[-1,1]$ and requantised before decoding to audio $\hat y=D_{\mathrm{SQ}}(Q(\operatorname{clip}(\widehat{\mathbf Z},-1,1)))$.

\noindent\textbf{Masking.} Two binary mask channels $\mathbf B$ indicate the target region and contextual MIDI. The region channel $\mathbf b_{\mathrm{obs}}$ is 1 on observed frames and 0 on target frames, giving the fixed audio condition $\mathbf A$:
\begin{equation}
\mathbf Z=Q(E_{\mathrm{SQ}}(y)),\qquad
\mathbf A=\mathbf b_{\mathrm{obs}}\odot\mathbf Z.
\label{eq:sq_audio}
\end{equation}
Thus $\mathbf A$ contains only observed SQ codes and remains fixed, without added noise, during generation.

\subsection{Frame-Aligned MIDI Span}
\label{sec:midi_span}

MIDI Span encodes notes from target and contextual MIDI as frame-aligned event sets with real-valued within-frame timing, then pools each set into one conditioning vector independently of slot order. Each non-drum note $j$, spanning $t_j^{\mathrm{on}}$ to $t_j^{\mathrm{off}}$, is represented by one event $e_{j,k}$ in each frame $k$ that overlaps this interval. The event state is \stateMono{ONSET} in the onset frame, \stateMono{OFFSET} in the offset frame, and \stateMono{SUSTAIN} in between. If both boundaries fall in one frame, only an \stateMono{ONSET} event is used. Drum hits use a single \stateMono{ONSET} event with zero remaining duration.

\noindent\textbf{Attributes.} Each event has two categorical attributes: event state and instrument class (including drums). Its four real-valued attributes are pitch, velocity, boundary position, and remaining duration, each normalised to $[-1,1]$:
\begin{equation*}
e_{j,k}=\bigl(c_{\mathrm{state}},c_{\mathrm{instr}},x_{\mathrm{pitch}},x_{\mathrm{vel}},x_{\mathrm{boundary}},x_{\mathrm{rem}}\bigr).
\end{equation*}
For an onset or offset at audio time $t_{\mathrm{event}}$ in a frame starting at $t_k$, the boundary position is
\begin{equation}
x_{\mathrm{boundary}}=2(t_{\mathrm{event}}-t_k)/H-1.
\label{eq:span_timing}
\end{equation}
This linearly maps within-frame timing to a continuous value in $[-1,1]$. At shared frame boundaries, onsets belong to the later frame and offsets to the earlier one. \stateMono{SUSTAIN} events use $x_{\mathrm{boundary}}=0$. Pitch and velocity are linearly scaled. For non-drum notes, remaining duration is the time to offset from onset (\stateMono{ONSET}) or frame start (otherwise), clipped to 20.48~s and log-scaled.

\noindent\textbf{Event-set encoding.} Frame-$k$ events form an unordered set $E_k$, padded to 128 slots. A shared encoder maps events to 128-dimensional vectors. Following Deep Sets~\cite{zaheer2017deep}, we pool independently of slot order by concatenating one sum and five learned sigmoid-gated sums, each root-mean-square (RMS) normalised. Because RMS-norm can suppress magnitude differences reflecting event counts, we add projected, log-scaled counts of melodic/drum events from $c_{\mathrm{instr}}$ and \stateMono{ONSET}/\stateMono{SUSTAIN}/\stateMono{OFFSET} events from $c_{\mathrm{state}}$, yielding $m_k\in\mathbb R^{768}$. Padding is masked and slot-index embeddings are omitted. Empty sets, including context frames without MIDI, yield zeros. Stacking $m_k$ in audio-frame order gives the MIDI condition $\mathbf M$.

\vspace{-6pt}
\subsection{Conditional Flow Matching}
\label{sec:conditional_flow}

\noindent\textbf{Training.} Conditional flow matching~\cite{lipman2023flow} mixes the clean SQ sequence $\mathbf Z$ with Gaussian noise $\boldsymbol\epsilon\sim\mathcal N(0,I)$ at flow time $t\sim\mathcal U[0,1]$. The noisy input $\mathbf Z_t$ and target vector field $\mathbf V^\star$ are
\begin{equation}
\mathbf Z_t=(1-t)\boldsymbol\epsilon+t\mathbf Z,
\qquad \mathbf V^\star=\mathbf Z-\boldsymbol\epsilon.
\label{eq:flow_path}
\end{equation}
Observed and target frames share this path. Observed audio enters through both noisy $\mathbf Z_t$ and fixed, clean $\mathbf A$ (Fig.~\ref{fig:conditioning}(a)).

Our generator (480.8M parameters, codec excluded) uses a 25-block diffusion transformer (DiT, width 1,024) $v_\theta$~\cite{peebles2023scalable} to predict the vector field $\widehat{\mathbf V}_t$ from noisy $\mathbf Z_t$, conditioned on $\mathbf A$, $\mathbf M$, and $\mathbf B$:
\begin{equation}
\widehat{\mathbf V}_t=v_\theta(\mathbf Z_t,t;\mathbf A,\mathbf M,\mathbf B).
\label{eq:conditional_velocity}
\end{equation}
At the DiT input, we add separate projections of $[\mathbf Z_t\Vert\mathbf M\Vert\mathbf B]$ and $\mathbf A$, where $\Vert$ denotes feature-wise concatenation. Padding is masked, and mean squared error between $\widehat{\mathbf V}_t$ and $\mathbf V^\star$ is minimised on target frames.

\noindent\textbf{Inference.} Euler integration starts from noise at $t=0$ and reaches the final latent sequence $\widehat{\mathbf Z}$ at $t=1$, with $\mathbf A$, $\mathbf M$, and $\mathbf B$ held fixed. At each step, observed frames follow their known interpolation in Eq.~\eqref{eq:flow_path}, recovering their original SQ codes at $t=1$. Intermediate flow states are not quantised, and only the final sequence is decoded as described in Sec.~\ref{sec:audio_sq}.

\setcounter{dbltopnumber}{1}
\begin{table*}[t]
\centering
\caption{Synthesis performance. Bold: best per group. MSS $\times10^3$. F-scores in \%. \textemdash: $F_{\mathrm{P37}}$/$F_{\mathrm{P13}}$ omitted for single-instrument data.}
\label{tab:synth_trial}
\begingroup
\fontsize{7.5}{8.5}\selectfont
\setlength{\tabcolsep}{1pt}
\begin{tabular*}{\textwidth}{@{\extracolsep{\fill}}ll*{10}{c}@{}}
\hline
\noalign{\vskip 2pt}
 & & \multicolumn{2}{c}{\textbf{Audio Quality}} & \multicolumn{4}{c}{\textbf{Audio Similarity}} & \multicolumn{4}{c}{\textbf{Note Adherence}} \\[2pt]
\cline{3-4}\cline{5-8}\cline{9-12}
\noalign{\vskip 3pt}
Dataset & Model & \metricHead{MuQ\metricSub{eval}$\uparrow$} & \metricHead{PQ$\uparrow$} & \metricHead{FAD\metricSub{MERT}$\downarrow$} & \metricHead{MuQ\metricSub{cos}$\uparrow$} & \metricHead{CLAP\metricSub{cos}$\uparrow$} & \metricHead{MSS$\downarrow$} & \metricHead{\textit{F}\metricSub{On}$\uparrow$} & \metricHead{\textit{F}\metricSub{P37}$\uparrow$} & \metricHead{\textit{F}\metricSub{P13}$\uparrow$} & \metricHead{\textit{F}\metricSub{OpenMIC}$\uparrow$} \\[2pt]
\hline
\noalign{\vskip 2pt}
\textbf{Slakh} & CTD$^{\dagger}$ & 4.20 & 7.49 & 2.00 & 0.81 & 0.78 & 36.46 & 46.91 & \textbf{33.31} & \textbf{38.36} & 12.25 \\
 & TokenSynth$^{\dagger}$ & 4.10 & 7.27 & 4.63 & 0.73 & 0.51 & 129.45 & 42.13 & 24.46 & 31.34 & 16.42 \\
 & Ours & \textbf{4.67} & \textbf{8.02} & \textbf{1.41} & \textbf{0.89} & \textbf{0.86} & \textbf{12.11} & \textbf{50.88} & 29.57 & 37.61 & \textbf{19.30} \\
\noalign{\vskip 1pt\hbox{\textcolor[gray]{0.65}{\rule{\textwidth}{0.3pt}}}\vskip 1pt}
\textbf{+drums} & SpecDiff & 4.51 & 7.80 & 4.85 & 0.78 & 0.64 & 49.03 & 51.84 & \textbf{39.64} & 43.12 & \textbf{9.63} \\
 & U-MusT & 3.75 & 7.95 & 4.16 & 0.63 & 0.73 & 49.02 & 23.17 & 11.00 & 13.75 & 5.89 \\
 & Ours & \textbf{4.65} & \textbf{8.08} & \textbf{1.33} & \textbf{0.89} & \textbf{0.90} & \textbf{29.42} & \textbf{54.06} & 37.78 & \textbf{44.84} & 9.51 \\
\hline
\noalign{\vskip 2pt}
\textbf{MusicNet} & SpecDiff & 4.38 & 6.87 & 3.50 & 0.90 & 0.75 & 25.21 & 68.25 & 66.79 & 67.24 & 71.67 \\
 & U-MusT & 4.16 & 7.09 & 4.03 & 0.84 & 0.84 & 30.84 & 44.31 & 41.19 & 42.53 & \textbf{75.00} \\
 & Ours & \textbf{4.75} & \textbf{7.60} & \textbf{3.15} & \textbf{0.92} & \textbf{0.87} & \textbf{22.02} & \textbf{76.56} & \textbf{71.94} & \textbf{74.69} & 68.33 \\
\hline
\noalign{\vskip 2pt}
\textbf{URMP} & SpecDiff & 4.58 & 7.52 & 6.53 & 0.89 & 0.67 & 17.23 & 50.50 & 34.62 & 44.80 & \textbf{69.19} \\
 & CTD$^{\dagger}$ & 3.91 & 7.13 & 10.89 & 0.77 & 0.55 & 77.02 & 35.39 & 2.89 & 22.56 & 6.67 \\
 & Ours & \textbf{4.82} & \textbf{7.90} & \textbf{4.95} & \textbf{0.91} & \textbf{0.83} & \textbf{9.50} & \textbf{54.24} & \textbf{37.85} & \textbf{47.90} & 53.00 \\
\hline
\noalign{\vskip 2pt}
\textbf{GuitarSet} & SpecDiff & 4.06 & 7.89 & 5.92 & 0.89 & 0.71 & 17.76 & 83.67 & \textemdash & \textemdash & 80.83 \\
 & CTD & 3.56 & 7.95 & 8.67 & 0.76 & 0.57 & 48.41 & 47.37 & \textemdash & \textemdash & 20.00 \\
 & TokenSynth & 3.70 & 7.33 & 14.77 & 0.69 & 0.37 & 110.17 & 34.36 & \textemdash & \textemdash & 11.67 \\
 & Ours & \textbf{4.14} & \textbf{8.40} & \textbf{4.05} & \textbf{0.92} & \textbf{0.91} & \textbf{14.29} & \textbf{88.01} & \textemdash & \textemdash & \textbf{85.00} \\
\hline
\noalign{\vskip 2pt}
\smash{\begin{tabular}[t]{@{}l@{}}\textbf{MAESTRO}\\Piano V3\end{tabular}} & SpecDiff & 4.47 & 6.58 & 4.46 & 0.93 & 0.71 & 31.24 & 66.63 & \textemdash & \textemdash & \textbf{100.00} \\
 & CTD & 3.75 & 6.89 & 6.82 & 0.74 & 0.66 & 49.22 & 16.67 & \textemdash & \textemdash & 96.67 \\
 & TokenSynth & 3.95 & 6.99 & 9.00 & 0.79 & 0.47 & 63.15 & 22.90 & \textemdash & \textemdash & 60.00 \\
 & MIDI-VALLE & 4.37 & \textbf{7.64} & 3.67 & 0.93 & \textbf{0.86} & 23.56 & 58.27 & \textemdash & \textemdash & \textbf{100.00} \\
 & U-MusT & 4.35 & 7.24 & \textbf{2.59} & 0.91 & \textbf{0.86} & 30.58 & 36.73 & \textemdash & \textemdash & \textbf{100.00} \\
 & Ours & \textbf{4.61} & 7.50 & 2.89 & \textbf{0.96} & \textbf{0.86} & \textbf{20.23} & \textbf{76.49} & \textemdash & \textemdash & \textbf{100.00} \\
\hline
\end{tabular*}
\endgroup
\end{table*}

\vspace{2pt}
\noindent\begin{minipage}[t]{\columnwidth}
\noindent\textbf{Optional FlowEdit.} We adapt FlowEdit~\cite{kulikov2025flowedit} from text-guided image editing to MIDI-guided audio editing without retraining. The edited latent sequence starts from before-edit SQ codes. Each step mixes these codes with fresh noise for the first DiT input. The second adds the difference between the current edited sequence and the before-edit codes to this noisy input. Both share $\mathbf A$ but use before-edit and revised MIDI, respectively. An Euler step subtracts the first predicted vector field from the second to update only target frames, preserving observed SQ codes.
\end{minipage}\par

\setlength{\dbltextfloatsep}{4pt plus 2pt minus 2pt}
\setlength{\textfloatsep}{4pt plus 2pt minus 2pt}
\begin{table*}[t]
\centering
\caption{Editing performance. $R$- and F-scores in \%. \textemdash: no requested deletions. Other notation follows Table~\ref{tab:synth_trial}.}
\label{tab:edit_trial}
\begingroup
\fontsize{7.5}{8.5}\selectfont
\setlength{\tabcolsep}{1pt}
\begin{tabular*}{\textwidth}{@{\extracolsep{\fill}}ll*{10}{r}@{}}
\hline
\noalign{\vskip 2pt}
 & & \multicolumn{2}{c}{\textbf{Audio Quality}} & \multicolumn{4}{c}{\textbf{Audio Similarity}} & \multicolumn{4}{c}{\textbf{Note Adherence}} \\[2pt]
\cline{3-4}\cline{5-8}\cline{9-12}
\noalign{\vskip 3pt}
Dataset / Task & Model & \metricHead{MuQ\metricSub{eval}$\uparrow$} & \metricHead{PQ$\uparrow$} & \metricHead{FAD\metricSub{MERT}$\downarrow$} & \metricHead{MuQ\metricSub{cos}$\uparrow$} & \metricHead{CLAP\metricSub{cos}$\uparrow$} & \metricHead{MSS$\downarrow$} & \metricHead{\textit{R}\metricSub{add}$\uparrow$} & \metricHead{\textit{R}\metricSub{del}$\downarrow$} & \metricHead{\textit{F}\metricSub{keep}$\uparrow$} & \metricHead{\textit{F}\metricSub{OpenMIC}$\uparrow$} \\[2pt]
\hline
\noalign{\vskip 2pt}
\smash{\begin{tabular}[t]{@{}l@{}}\textbf{Slakh (+drums)}\\Add new\\instrument\end{tabular}} & SpecDiff & 4.29 & 7.71 & 4.87 & 0.79 & 0.63 & 45.04 & 43.68 & \textemdash & 32.54 & 12.86 \\
 & U-MusT & 3.72 & 7.98 & 3.45 & 0.70 & 0.76 & 41.39 & 21.32 & \textemdash & 8.34 & 11.66 \\
 & Ours & \textbf{4.44} & \textbf{8.13} & 1.68 & 0.87 & \textbf{0.84} & 31.54 & \textbf{49.13} & \textemdash & 31.09 & 14.73 \\
 & Ours + FlowEdit & 4.34 & 8.01 & \textbf{1.43} & \textbf{0.88} & 0.83 & \textbf{24.40} & 42.14 & \textemdash & \textbf{32.68} & \textbf{14.85} \\
\hline
\noalign{\vskip 2pt}
\smash{\begin{tabular}[t]{@{}l@{}}\textbf{Slakh}\\Note ins/del\end{tabular}} & CTD$^{\dagger}$ & 3.92 & 7.28 & 1.82 & 0.82 & 0.72 & 80.55 & 43.20 & \textbf{3.82} & 28.91 & 17.33 \\
 & Ours & \textbf{4.40} & \textbf{7.88} & 1.25 & 0.90 & 0.84 & 21.37 & 42.51 & 6.08 & 33.24 & \textbf{26.56} \\
 & Ours + FlowEdit & 4.32 & 7.76 & \textbf{1.06} & \textbf{0.91} & \textbf{0.86} & \textbf{11.15} & \textbf{43.49} & 8.01 & \textbf{35.75} & 24.56 \\
\noalign{\vskip 1pt\hbox{\textcolor[gray]{0.65}{\rule{\textwidth}{0.3pt}}}\vskip 1pt}
\textbf{+drums} & SpecDiff & 4.25 & 7.72 & 4.15 & 0.80 & 0.62 & 40.33 & 52.64 & \textbf{2.62} & 39.19 & 15.48 \\
 & U-MusT & 3.72 & 7.95 & 2.65 & 0.70 & 0.74 & 37.81 & 20.50 & 3.72 & 9.56 & 13.28 \\
 & Ours & \textbf{4.44} & \textbf{8.03} & 1.03 & 0.89 & 0.87 & 23.26 & \textbf{56.44} & 4.56 & \textbf{41.90} & \textbf{16.92} \\
 & Ours + FlowEdit & 4.36 & 7.93 & \textbf{0.83} & \textbf{0.91} & \textbf{0.88} & \textbf{12.65} & 50.54 & 7.78 & 41.55 & 16.21 \\
\hline
\noalign{\vskip 2pt}
\smash{\begin{tabular}[t]{@{}l@{}}\textbf{POP909 (piano)}\\AI edits~\cite{bhandari2026change}\end{tabular}} & SpecDiff & 4.35 & 7.47 & 5.46 & 0.87 & 0.76 & 30.92 & 83.03 & \textbf{2.60} & 12.85 & 98.85 \\
 & CTD & 3.66 & 7.42 & 7.99 & 0.72 & 0.65 & 45.11 & 30.16 & 6.74 & 3.23 & 97.03 \\
 & TokenSynth & 3.79 & 7.41 & 9.21 & 0.74 & 0.63 & 87.71 & 27.24 & 5.27 & 7.66 & 80.50 \\
 & MIDI-VALLE & 4.13 & 7.95 & 5.19 & 0.86 & 0.73 & 23.93 & 72.74 & 4.86 & 10.41 & 99.74 \\
 & U-MusT & 4.26 & 7.78 & 6.50 & 0.84 & \textbf{0.81} & 29.00 & 42.56 & 5.39 & 3.77 & \textbf{99.96} \\
 & Ours & \textbf{4.42} & 7.99 & 3.33 & \textbf{0.92} & 0.77 & 16.82 & \textbf{89.86} & 3.40 & \textbf{13.84} & 99.81 \\
 & Ours + FlowEdit & 4.36 & \textbf{8.03} & \textbf{2.31} & 0.91 & 0.78 & \textbf{15.64} & 86.81 & 5.06 & 12.46 & 99.89 \\
\hline
\end{tabular*}
\endgroup
\end{table*}

\begin{table}[!t]
\makeatletter
\long\def\@makecaption#1#2{%
  \vskip 4pt
  \setbox\@tempboxa\hbox{#1. #2}%
  \ifdim\wd\@tempboxa>\hsize #1. #2\par
  \else\hbox to\hsize{\hfil\box\@tempboxa\hfil}\fi}
\makeatother
\centering
\caption{Tracing note-adherence error sources on Slakh (+drums). Scores in \%; $\Delta$: drop from the preceding row (percentage points). \textcolor{red}{$\blacktriangledown$}: $\ge10$; \textcolor{red}{$\triangledown$}: $<10$. AMT: automatic music transcription.}
\label{tab:codec_edit_controls}
\begingroup
\fontsize{7.5}{8.5}\selectfont
\setlength{\tabcolsep}{1pt}
\newcommand{\scoreDrop}[2]{\makebox[18pt][r]{#1}\,\makebox[30pt][l]{(\textcolor{red}{$\ifdim #2pt<10pt\triangledown\else\blacktriangledown\fi$}\,\makebox[18pt][r]{#2})}}
\newcommand{\scoreOnly}[1]{\makebox[18pt][r]{#1}\,\makebox[30pt][l]{}}
\newcommand{\scoreRow}[1]{\hspace{4pt}\makebox[7pt][l]{$-$}#1}
\begin{tabular*}{\columnwidth}{@{\extracolsep{\fill}}p{127pt}rr@{}}
\hline
\noalign{\vskip 2pt}
\textbf{(a) Slakh / Synthesis} & \multicolumn{1}{c}{\metricHead{\textit{F}\metricSub{P37}$\uparrow$ ($\Delta$)}} & \multicolumn{1}{c}{\metricHead{\textit{F}\metricSub{On}$\uparrow$ ($\Delta$)}} \\[2pt]
\hline
\noalign{\vskip 2pt}
Ground Truth (GT) w/ Perfect AMT & \scoreOnly{100} & \scoreOnly{100} \\
\scoreRow{GT w/ AMT} & \scoreDrop{78.40}{21.60} & \scoreDrop{82.79}{17.21} \\
\scoreRow{Reconstruction (no SQ) w/ AMT} & \scoreDrop{44.43}{33.97} & \scoreDrop{55.94}{26.85} \\
\scoreRow{Reconstruction (SQ) w/ AMT} & \scoreDrop{43.98}{0.45} & \scoreDrop{55.79}{0.15} \\
\scoreRow{Generated (ours)} & \scoreDrop{37.78}{6.20} & \scoreDrop{54.06}{1.73} \\[2pt]
\hline
\end{tabular*}

\vspace{2pt}
\begin{tabular*}{\columnwidth}{@{\extracolsep{\fill}}p{127pt}rr@{}}
\textbf{(b) Slakh / Add new instrument} & \multicolumn{1}{c}{\metricHead{\textit{R}\metricSub{add}$\uparrow$ ($\Delta$)}} & \multicolumn{1}{c}{\metricHead{\textit{F}\metricSub{keep}$\uparrow$ ($\Delta$)}} \\[2pt]
\hline
\noalign{\vskip 2pt}
Ground Truth (GT) w/ Perfect AMT & \scoreOnly{100} & \scoreOnly{100} \\
\scoreRow{GT w/ AMT} & \scoreDrop{71.40}{28.60} & \scoreDrop{72.80}{27.20} \\
\scoreRow{Reconstruction (no SQ) w/ AMT} & \scoreDrop{55.99}{15.41} & \scoreDrop{39.01}{33.79} \\
\scoreRow{Reconstruction (SQ) w/ AMT} & \scoreDrop{54.54}{1.45} & \scoreDrop{37.50}{1.51} \\
\scoreRow{Generated (ours)} & \scoreDrop{49.13}{5.41} & \scoreDrop{31.09}{6.41} \\[2pt]
\hline
\end{tabular*}
\endgroup

\par
\centering
\caption{Inference-time ablations. CM: contextual MIDI. CA: clean audio condition $\mathbf A$. Bold: best per group.}
\label{tab:euler_steps}
\begingroup
\fontsize{7.5}{8.5}\selectfont
\setlength{\tabcolsep}{1pt}
\begin{tabular*}{\columnwidth}{@{\extracolsep{\fill}}lrrrrrrrrr@{}}
\hline
\noalign{\vskip 2pt}
 & \multicolumn{5}{c}{Euler steps (Base)} & \multicolumn{4}{c}{32-step ablations} \\
\noalign{\vskip 1pt}
\omit & \multispan{5}\hskip 2pt\leaders\hrule height 0.4pt\hfill\hskip 2pt
 & \multispan{4}\hskip 2pt\leaders\hrule height 0.4pt\hfill\hskip 2pt\cr
\noalign{\vskip 3pt}
 & 4 & 8 & 16 & \textbf{32} & 64 & Base & $-$CM & $-$CA & $-$CFG \\[2pt]
\hline
\noalign{\vskip 2pt}
\metricHead{MuQ\metricSub{eval}$\uparrow$} & \textbf{4.48} & 4.46 & 4.43 & 4.41 & 4.40 & 4.41 & \textbf{4.44} & 3.48 & 4.29 \\
\metricHead{FAD\metricSub{MERT}$\downarrow$} & 3.65 & 3.29 & 3.09 & 2.99 & \textbf{2.95} & 2.99 & 3.36 & 9.95 & \textbf{2.98} \\
\metricHead{\textit{F}\metricSub{On} (\%)$\uparrow$} & 74.60 & \textbf{76.14} & 75.90 & 73.34 & 73.03 & 73.34 & \textbf{76.01} & 45.66 & 65.67 \\
\hline
\end{tabular*}
\endgroup

\end{table}

\begingroup
\setlength{\dbltextfloatsep}{8pt plus 2pt minus 2pt}
\setlength{\parskip}{2pt}
\section{Experiments}
\label{sec:experiments}

\subsection{Experimental Setup}
\label{sec:experimental_setup}

\noindent\textbf{Tasks and data.} For synthesis, we test whether generated audio follows target MIDI and joins surrounding audio smoothly with matching timbres and acoustics (Table~\ref{tab:synth_trial}). For editing, we test preservation of original timbres and unchanged notes under revised MIDI (Table~\ref{tab:edit_trial}). Each edit provides paired MIDI and ground-truth audio before and after revision. Adding or removing stems or stem segments from Slakh test songs~\cite{manilow2019slakh} yields 668 original--revised pairs. When adding a new instrument, the reference audio excludes it. Rendering original and Modulator-revised MIDI~\cite{bhandari2026change} for 147 POP909 songs~\cite{wang2020pop909} with the Salamander piano SoundFont~\cite{holm2020salamander} yields 882 such pairs, half dry and half with matched reverberation. Context is unchanged and excluded from evaluation.

\noindent\textbf{Training and inference.} We train on 660~h of performance MIDI paired with 48-kHz mono audio from 17 public instrumental datasets (15 with real recordings). We use Adam for 125k steps (effective batch 80, peak learning rate $10^{-4}$, linear warmup and decay). In Tables~\ref{tab:synth_trial}--\ref{tab:edit_trial}, our model uses contextual audio and MIDI, 32 Euler steps, and classifier-free guidance (CFG~2). On a GH200 (BF16, batch 1), MIDI-only generation with SQ decoding takes 0.93~s per 20.48~s of audio, with 3.37~GiB peak reserved memory.

\noindent\textbf{Baselines.} We prioritise baselines trained on the corresponding dataset and use official checkpoints. For POP909, all models, including ours, were trained on piano audio but not the Salamander timbre. \emph{For multi-instrument results ($\dagger$), we give CTD~\cite{demerle2024combining} and TokenSynth~\cite{kim2025tokensynth} isolated stem context and mix their generated stems at fixed gains without rebalancing.} \mbox{SpecDiff}~\cite{hawthorne2022spectrogramdiffusion} and U-MusT~\cite{jung2026umust} generate mixtures, while MIDI-VALLE~\cite{tang2025midivalle} is piano-only. All three use only preceding audio, with contextual MIDI only for U-MusT and MIDI-VALLE.

\noindent\textbf{Metrics.} \emph{Audio Quality} evaluates generated audio alone using MuQ-Eval's musical impression score~\cite{zhu2026muqeval} and Audiobox production quality (PQ)~\cite{tjandra2025audiobox}. \emph{Audio Similarity} compares generated and ground-truth audio using MERT-based FAD~\cite{kilgour2019fad,li2024mert}, MuQ/CLAP cosine similarity~\cite{zhu2025muq,wu2023clap}, and Smooth MSS spectral error~\cite{schwaer2023multiscale}. \emph{\mbox{Note Adherence}} compares YourMT3+~\cite{chang2024yourmt3} transcriptions of generated audio with target MIDI at a 50-ms onset tolerance. $F_{\mathrm{On}}$ is instrument-agnostic note onset F1, while Multi Onset F1 also matches instrument labels using 37-class fine ($F_{\mathrm{P37}}$) and 13-class coarse ($F_{\mathrm{P13}}$) vocabularies~\cite{chang2024yourmt3}. For editing, 37-class matching measures recall of requested additions ($R_{\mathrm{add}}$) and F1 of unchanged notes ($F_{\mathrm{keep}}$). For $F_{\mathrm{keep}}$, precision excludes predictions matched to requested additions or deletions. $R_{\mathrm{del}}$ is the fraction of notes requested for deletion that remain detected. $F_{\mathrm{OpenMIC}}$ measures 20-class instrument-presence F1~\cite{koutini2022passt}. Our repository will include dataset references and 95\% bootstrap confidence intervals for per-example metrics.

\subsection{Synthesis and Editing}
\label{sec:paired_evaluation}

\noindent\textbf{Synthesis.} SpanSynth-Edit (denoted Ours in Tables~\ref{tab:synth_trial}--\ref{tab:edit_trial}) matches or outperforms the reported baselines on the audio-quality and audio-similarity metrics in Table~\ref{tab:synth_trial}, except for PQ and FAD on MAESTRO. There, MIDI-VALLE, a piano-only model trained on a large piano dataset~\cite{tang2025midivalle}, leads in PQ, while U-MusT leads in FAD.
In each Table~\ref{tab:synth_trial} group, our model's $F_{\mathrm{On}}$ exceeds the highest baseline score by 2.22--9.86 percentage points, but this advantage is inconsistent for scores requiring instrument-label matches. On Slakh (+drums), our model exceeds SpecDiff in $F_{\mathrm{P13}}$ but falls behind in $F_{\mathrm{P37}}$.

\noindent\textbf{Editing.} Our model without FlowEdit outperforms the reported baselines in MuQ-Eval and PQ in every Table~\ref{tab:edit_trial} group. Applying FlowEdit (denoted Ours + FlowEdit) yields the best FAD and MSS among the compared systems in each group but reduces MuQ-Eval relative to our model without FlowEdit.

The highest $R_{\mathrm{add}}$ and $F_{\mathrm{keep}}$ scores in each Table~\ref{tab:edit_trial} group come from our model with or without FlowEdit. However, our models obtain worse $R_{\mathrm{del}}$ scores than CTD on non-drum Slakh and SpecDiff on Slakh (+drums) and POP909. Relative to Ours, Ours + FlowEdit increases $R_{\mathrm{del}}$ in every Table~\ref{tab:edit_trial} deletion group and decreases $R_{\mathrm{add}}$ in three of four groups. Its effect on $F_{\mathrm{keep}}$ varies by task.

Note-adherence scores assess whether audio follows MIDI but also reflect transcription errors, especially in multi-instrument mixtures.
\par\endgroup

\newpage
\subsection{Analysis}
\label{sec:onset_control}
\label{sec:efficiency}

\noindent\textbf{Note-adherence errors.} In Table~\ref{tab:codec_edit_controls}(a,b), we compare ground-truth, reconstructed, and generated audio to break down the main sources of error. We observe substantial note-adherence score losses  originating from the YourMT3+ transcription model itself (\(F_{\mathrm{P37}}\): 21.60, \(R_{\mathrm{add}}\): 28.60). This is especially pronounced in mixtures with many instruments. Additional losses arise from encoder-decoder reconstruction without SQ (\(F_{\mathrm{P37}}\): 33.97, \(R_{\mathrm{add}}\): 15.41). Quantisation adds only small losses (\(F_{\mathrm{P37}}\): 0.45, \(R_{\mathrm{add}}\): 1.45). Transcription and reconstruction without SQ account for most losses across all four metrics, with smaller losses from generation. Reconstruction losses may reflect a distribution mismatch between HeartCodec's pretraining data and ours.

\begin{figure}[!t]
\makeatletter
\long\def\@makecaption#1#2{%
  \vskip 4pt
  \setbox\@tempboxa\hbox{#1. #2}%
  \ifdim\wd\@tempboxa>\hsize #1. #2\par
  \else\hbox to\hsize{\hfil\box\@tempboxa\hfil}\fi}
\makeatother
\centering
\raisebox{0pt}[148pt][0pt]{\includegraphics[width=244pt,height=148pt]{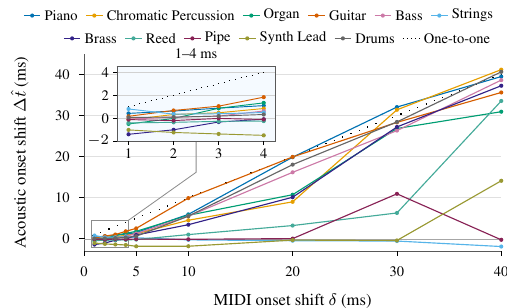}}
\caption{Median acoustic responses to 1--40~ms MIDI onset shifts. Dotted lines: equal shifts. Zoom: 1--4~ms.}
\label{fig:onset_control}
\end{figure}

\noindent\textbf{Timing sensitivity.}
We test the timing precision sensitivity of the synthesised audio to the MIDI span encoder by delaying MIDI onsets by 1--40 ms. We quantify the deviations between the synthesised reference and time-shifted re-synthesised audio via a cross-correlation of pitch harmonic energy (or drum-onset curves~\cite{bello2005onset} with peak interpolation). With a perfect MIDI encoder and synthesis model, a 5 ms onset shift, for example, will reflect a 5 ms shift in the output, representing the dotted diagonal in Figure~\ref{fig:onset_control}. In Fig.~\ref{fig:onset_control}, although the model is not sensitive to small onset shifts in the range of 1--5 ms, possibly due to annotation errors, it starts becoming receptive at noticeable shifts from 10--40 ms. Strings and Pipe show weak measured responses, possibly reflecting difficulties in localising non-percussive onsets~\cite{bello2005onset}.


\noindent\textbf{Ablations.} Table~\ref{tab:euler_steps} reveals a trade-off beyond 8 Euler steps: FAD decreases while onset F1 and MuQ-Eval decline. At 32 steps, contextual MIDI provides no consistent benefit. Removing clean contextual audio (CA; $\mathbf A$) severely degrades all three metrics. These results strongly support our design choice of conditioning on clean contextual audio latents (Sec.~\ref{sec:audio_sq}). Removing CFG also reduces MuQ-Eval and onset F1 with little FAD change.

\begin{samepage}
\section{Conclusion}
\label{sec:conclusion}

SpanSynth-Edit generates and edits multi-instrument mixtures with SQ latents and MIDI Span. On synthesis and editing benchmarks, our model outperforms the evaluated baselines on most objective audio-quality and audio-similarity metrics, while achieving comparable note adherence. We analysed the origin of note-adherence errors, timing precision of MIDI span to the generated output, and support for contextual audio conditioning.
Codec limitations motivate future decoder fine-tuning. We leave listening tests with music experts for future work. Audio examples are on our project page\footref{fn:project}.
\par\end{samepage}

\clearpage
\section{Acknowledgement}
Keshav Bhandari was supported by the UKRI Centre for Doctoral Training in Artificial Intelligence and Music (EP/S022694/1). EmotionWave provided GPU resources for part of the model training. We thank TaeGyun Kwon and Dabin Kim for their helpful feedback.

\AddToHookNext{cmd/thebibliography/after}{\setlength{\itemsep}{1pt}}
\bibliographystyle{IEEEbib}
\bibliography{refs}

\clearpage
\onecolumn
\appendix
\pagestyle{plain}
\fontsize{9}{11}\selectfont
\setlength{\parskip}{4pt}

\begin{bibunit}[IEEEbib]
\makeatletter
\def\@extra@b@citeb{.appendix}
\def\bibcite#1#2{%
  \global\@namedef{b@#1}{\hyper@@link[cite]{}{cite.#1.appendix}{A#2}}}
\makeatother

\section{Dataset List, References, and Splits}
\label{app:datasets}
We train on the 17 instrumental dataset families below, using instrumental stems where applicable. MusicNet denotes MusicNetEM, with revised note annotations for the original MusicNet recordings.

\begin{center}
\renewcommand{\arraystretch}{1.4}
\begin{tabular*}{\textwidth}{@{}r@{\hspace{12pt}}l@{\extracolsep{\fill}}l@{}}
\hline
 & Dataset & Audio content \\
\hline
1 & Slakh~\cite{manilow2019slakh} & Synthesised multi-instrument mixtures, including drums \\
2 & AAM~\cite{ostermann2023aam} & Artificial audio multitracks \\
3 & MAESTRO~\cite{hawthorne2019maestro} & Solo piano recordings, mostly classical \\
4 & PianoVAM~\cite{kim2025pianovam} & Piano performances \\
5 & MusicNetEM~\cite{thickstun2017musicnet,maman2022unaligned} & Classical instrumental recordings \\
6 & BSD / BSED~\cite{berendes2026bsd} & Beethoven symphony recordings / evaluation excerpts \\
7 & GOAT~\cite{loth2025goat} & Electric guitar recordings with clean and distorted tones \\
8 & GuitarSet~\cite{xi2018guitarset} & Acoustic guitar: jazz, rock, funk, bossa nova, singer-songwriter \\
9 & URMP~\cite{li2018urmp} & Classical ensembles of 2--5 separately recorded parts \\
10 & IDMT-SMT-Bass~\cite{abesser2010bass} & Electric bass recordings \\
11 & KRAISLER~\cite{kim2026kraisler} & Piano and violin duet recordings \\
12 & ChoraleBricks~\cite{balke2025choralebricks} & Wind-instrument multitracks \\
13 & FiloBass~\cite{riley2023filobass} & Jazz bass recordings \\
14 & MDB Drums~\cite{southall2017mdbdrums} & Annotated drum stems \\
15 & EGSet12~\cite{pedroza2024egset12} & Solo electric guitar performances \\
16 & ENST-Drums~\cite{gillet2006enst} & Drum performances \\
17 & STAR Drums~\cite{weber2025stardrums} & Drum transcription data \\
\hline
\end{tabular*}
\end{center}

\noindent\textbf{Official partitions.} Slakh and MAESTRO combine their official training and validation sets for training. PianoVAM uses its training and extended-training sets, while GOAT and STAR Drums use their official training sets without adding validation data. Their official test sets are reserved for evaluation. MDB Drums follows the published MIREX partition of 12 training and 11 test songs.

\noindent\textbf{Adopted partitions.} MusicNetEM uses the extended ten-recording test set rather than the original three-recording test set. These ten recordings are excluded from both acoustic and synthesised training versions. URMP follows the MT3 split~\cite{gardner2022mt3}, with 35 training and nine test pieces. GuitarSet uses progressions 1 and 2 for training and progression 3 for evaluation. ENST-Drums uses drummers 1 and 2 for training and drummer 3 for evaluation. IDMT-SMT-Bass retains the stored approximately 80:20 training/validation partition, with the latter reserved for evaluation.

\noindent\textbf{Custom partitions.} AAM, ChoraleBricks, FiloBass, and EGSet12 use our approximately 8:1:1 training/validation/test splits, assigned at the song or recording level. Only the training partition is included in the training pool. KRAISLER holds out tracks 05, 10, 15, and 20 for evaluation, using their studio and hall mixtures. The remaining 16 tracks provide training stems from all three room conditions.

\noindent\textbf{BSD and BSED.} BSD provides the training recordings, while BSED provides evaluation excerpts. These collections share symphonies and score passages, so this is not a composition-disjoint partition. The prepared BSED evaluation selects recordings not used for training and supplements them with synthesised renditions.

For editing, POP909~\cite{wang2020pop909} supplies piano arrangements of pop songs. Original and Modulator-revised MIDI~\cite{bhandari2026change} are rendered with the Salamander piano SoundFont~\cite{holm2020salamander} (Sec.~\ref{sec:experimental_setup}).

\noindent\textbf{Audio preprocessing and SQ storage.} Audio is downmixed to mono and resampled to 48~kHz as needed, including upsampling 44.1-kHz sources. We extract 20.48-s crops at 5.12-s intervals (75\% overlap), adding a final end-aligned crop and zero-padding short recordings. The frozen HeartCodec encoder also receives 2~s of preceding audio, zero-padded when unavailable. Its first 50 latent frames are discarded, and the remaining $512\times128$ SQ codes are stored as signed 8-bit integers.

\noindent\textbf{Augmentation.} We precompute mixtures retaining the base arrangement with up to two additional recordings. Training requests zero, one, or two additions in 50\%, 35\%, and 15\% of samples. Conflicting instrument parts are excluded, and additions exceeding the 128-event frame budget are rejected. Eligible drum-free crops receive length-preserving Rubber Band R3 pitch shifts of $\pm1$ or $\pm2$ semitones, with matched MIDI transposition before SQ encoding. Training requests shifted crops 25\% of the time, falling back to unshifted crops when no eligible shifted version is available.

\clearpage
\section{Synthesis Results with Confidence Intervals (Table 1)}
\label{app:synthesis_ci}

\noindent\textbf{Reporting convention.} Each per-example metric is reported as mean [95\% CI lower bound, upper bound]. Intervals use a source-recording cluster bootstrap with 10,000 resamples and the 2.5th and 97.5th percentiles. Each resampled recording contributes all its evaluated excerpts, and the mean is taken over excerpts. These intervals quantify variation across the evaluation recordings, not across training runs, and are not tests of paired differences between models. FAD is a distributional score computed from pooled features, so only its point estimate is shown. MSS is multiplied by $10^3$, and all note and instrument scores are percentages. Cosine similarities are shown to three decimal places.

\noindent\textbf{Comparison conditions.} Models and conditioning follow Table~\ref{tab:synth_trial} and Sec.~\ref{sec:experimental_setup}. A dagger denotes separate stem generation followed by mixing. Each Slakh group has 100 excerpts from 100 recordings. MusicNetEM has 20 excerpts from 10 recordings, URMP has 20 from 9, and GuitarSet and MAESTRO each have 20 from 20. The small number of recordings limits the precision of the MusicNetEM and URMP intervals. The single-instrument tables omit $F_{\mathrm{P37}}$ and $F_{\mathrm{P13}}$, as in the main paper.
\subsection{Slakh (without drums)}
\begin{center}
\renewcommand{\arraystretch}{1.2}
\setlength{\tabcolsep}{3pt}
\begin{tabular*}{\textwidth}{@{\extracolsep{\fill}}lccc@{}}
\hline
Model & MuQ$_{\mathrm{eval}}\uparrow$ & PQ$\uparrow$ & FAD$_{\mathrm{MERT}}\downarrow$ \\
\hline
CTD$^\dagger$ & 4.20 [4.14, 4.27] & 7.49 [7.39, 7.59] & 2.00 \\
TokenSynth$^\dagger$ & 4.10 [4.03, 4.16] & 7.27 [7.18, 7.35] & 4.63 \\
Ours & 4.67 [4.60, 4.74] & 8.02 [7.96, 8.07] & 1.41 \\
\hline
\end{tabular*}
\end{center}
\begin{center}
\renewcommand{\arraystretch}{1.2}
\setlength{\tabcolsep}{3pt}
\begin{tabular*}{\textwidth}{@{\extracolsep{\fill}}lccc@{}}
\hline
Model & MuQ$_{\mathrm{cos}}\uparrow$ & CLAP$_{\mathrm{cos}}\uparrow$ & MSS$\downarrow$ \\
\hline
CTD$^\dagger$ & 0.813 [0.796, 0.829] & 0.778 [0.763, 0.792] & 36.46 [35.03, 37.94] \\
TokenSynth$^\dagger$ & 0.733 [0.716, 0.749] & 0.513 [0.495, 0.531] & 129.45 [123.31, 135.62] \\
Ours & 0.890 [0.880, 0.899] & 0.856 [0.839, 0.871] & 12.11 [11.49, 12.72] \\
\hline
\end{tabular*}
\end{center}
\begin{center}
\renewcommand{\arraystretch}{1.2}
\setlength{\tabcolsep}{3pt}
\begin{tabular*}{\textwidth}{@{\extracolsep{\fill}}lcccc@{}}
\hline
Model & $F_{\mathrm{On}}\uparrow$ & $F_{\mathrm{P37}}\uparrow$ & $F_{\mathrm{P13}}\uparrow$ & $F_{\mathrm{OpenMIC}}\uparrow$ \\
\hline
CTD$^\dagger$ & 46.91 [44.52, 49.23] & 33.31 [30.96, 35.69] & 38.36 [35.84, 40.81] & 12.25 [8.64, 15.99] \\
TokenSynth$^\dagger$ & 42.13 [39.96, 44.27] & 24.46 [22.38, 26.49] & 31.34 [29.22, 33.46] & 16.42 [12.39, 20.64] \\
Ours & 50.88 [48.45, 53.30] & 29.57 [26.80, 32.41] & 37.61 [34.98, 40.29] & 19.30 [15.09, 23.57] \\
\hline
\end{tabular*}
\end{center}

\subsection{Slakh (+drums)}
\begin{center}
\renewcommand{\arraystretch}{1.2}
\setlength{\tabcolsep}{3pt}
\begin{tabular*}{\textwidth}{@{\extracolsep{\fill}}lccc@{}}
\hline
Model & MuQ$_{\mathrm{eval}}\uparrow$ & PQ$\uparrow$ & FAD$_{\mathrm{MERT}}\downarrow$ \\
\hline
SpecDiff & 4.51 [4.44, 4.57] & 7.80 [7.74, 7.85] & 4.85 \\
U-MusT & 3.75 [3.67, 3.83] & 7.95 [7.89, 8.02] & 4.16 \\
Ours & 4.65 [4.59, 4.70] & 8.08 [8.04, 8.12] & 1.33 \\
\hline
\end{tabular*}
\end{center}
\begin{center}
\renewcommand{\arraystretch}{1.2}
\setlength{\tabcolsep}{3pt}
\begin{tabular*}{\textwidth}{@{\extracolsep{\fill}}lccc@{}}
\hline
Model & MuQ$_{\mathrm{cos}}\uparrow$ & CLAP$_{\mathrm{cos}}\uparrow$ & MSS$\downarrow$ \\
\hline
SpecDiff & 0.778 [0.764, 0.790] & 0.637 [0.617, 0.657] & 49.03 [45.87, 52.62] \\
U-MusT & 0.627 [0.600, 0.653] & 0.734 [0.712, 0.754] & 49.02 [46.04, 52.16] \\
Ours & 0.889 [0.880, 0.897] & 0.895 [0.883, 0.907] & 29.42 [28.13, 30.71] \\
\hline
\end{tabular*}
\end{center}
\begin{center}
\renewcommand{\arraystretch}{1.2}
\setlength{\tabcolsep}{3pt}
\begin{tabular*}{\textwidth}{@{\extracolsep{\fill}}lcccc@{}}
\hline
Model & $F_{\mathrm{On}}\uparrow$ & $F_{\mathrm{P37}}\uparrow$ & $F_{\mathrm{P13}}\uparrow$ & $F_{\mathrm{OpenMIC}}\uparrow$ \\
\hline
SpecDiff & 51.84 [49.18, 54.43] & 39.64 [36.62, 42.68] & 43.12 [40.24, 46.01] & 9.63 [7.00, 12.39] \\
U-MusT & 23.17 [21.52, 24.98] & 11.00 [9.57, 12.52] & 13.75 [12.20, 15.40] & 5.89 [3.83, 8.18] \\
Ours & 54.06 [51.59, 56.48] & 37.78 [35.17, 40.40] & 44.84 [42.19, 47.40] & 9.51 [6.88, 12.28] \\
\hline
\end{tabular*}
\end{center}

\clearpage

\subsection{MusicNetEM}
\begin{center}
\renewcommand{\arraystretch}{1.2}
\setlength{\tabcolsep}{3pt}
\begin{tabular*}{\textwidth}{@{\extracolsep{\fill}}lccc@{}}
\hline
Model & MuQ$_{\mathrm{eval}}\uparrow$ & PQ$\uparrow$ & FAD$_{\mathrm{MERT}}\downarrow$ \\
\hline
SpecDiff & 4.38 [4.18, 4.61] & 6.87 [6.43, 7.26] & 3.50 \\
U-MusT & 4.16 [3.85, 4.49] & 7.09 [6.76, 7.40] & 4.03 \\
Ours & 4.75 [4.48, 5.00] & 7.60 [7.30, 7.87] & 3.15 \\
\hline
\end{tabular*}
\end{center}
\begin{center}
\renewcommand{\arraystretch}{1.2}
\setlength{\tabcolsep}{3pt}
\begin{tabular*}{\textwidth}{@{\extracolsep{\fill}}lccc@{}}
\hline
Model & MuQ$_{\mathrm{cos}}\uparrow$ & CLAP$_{\mathrm{cos}}\uparrow$ & MSS$\downarrow$ \\
\hline
SpecDiff & 0.896 [0.873, 0.917] & 0.745 [0.696, 0.789] & 25.21 [18.54, 32.35] \\
U-MusT & 0.837 [0.797, 0.876] & 0.842 [0.794, 0.881] & 30.84 [21.69, 42.02] \\
Ours & 0.921 [0.906, 0.934] & 0.869 [0.830, 0.899] & 22.02 [16.21, 28.54] \\
\hline
\end{tabular*}
\end{center}
\begin{center}
\renewcommand{\arraystretch}{1.2}
\setlength{\tabcolsep}{3pt}
\begin{tabular*}{\textwidth}{@{\extracolsep{\fill}}lcccc@{}}
\hline
Model & $F_{\mathrm{On}}\uparrow$ & $F_{\mathrm{P37}}\uparrow$ & $F_{\mathrm{P13}}\uparrow$ & $F_{\mathrm{OpenMIC}}\uparrow$ \\
\hline
SpecDiff & 68.25 [57.59, 79.36] & 66.79 [55.45, 78.54] & 67.24 [56.21, 78.79] & 71.67 [46.67, 93.33] \\
U-MusT & 44.31 [36.06, 53.93] & 41.19 [31.90, 51.88] & 42.53 [33.54, 52.79] & 75.00 [48.33, 96.67] \\
Ours & 76.56 [69.55, 84.40] & 71.94 [62.55, 82.21] & 74.69 [66.49, 83.60] & 68.33 [43.33, 90.00] \\
\hline
\end{tabular*}
\end{center}

\subsection{URMP}
\begin{center}
\renewcommand{\arraystretch}{1.2}
\setlength{\tabcolsep}{3pt}
\begin{tabular*}{\textwidth}{@{\extracolsep{\fill}}lccc@{}}
\hline
Model & MuQ$_{\mathrm{eval}}\uparrow$ & PQ$\uparrow$ & FAD$_{\mathrm{MERT}}\downarrow$ \\
\hline
SpecDiff & 4.58 [4.48, 4.65] & 7.52 [7.25, 7.70] & 6.53 \\
CTD$^\dagger$ & 3.91 [3.71, 4.13] & 7.13 [6.98, 7.28] & 10.89 \\
Ours & 4.82 [4.67, 4.96] & 7.90 [7.67, 8.07] & 4.95 \\
\hline
\end{tabular*}
\end{center}
\begin{center}
\renewcommand{\arraystretch}{1.2}
\setlength{\tabcolsep}{3pt}
\begin{tabular*}{\textwidth}{@{\extracolsep{\fill}}lccc@{}}
\hline
Model & MuQ$_{\mathrm{cos}}\uparrow$ & CLAP$_{\mathrm{cos}}\uparrow$ & MSS$\downarrow$ \\
\hline
SpecDiff & 0.885 [0.865, 0.900] & 0.671 [0.647, 0.699] & 17.23 [12.13, 25.03] \\
CTD$^\dagger$ & 0.767 [0.748, 0.789] & 0.555 [0.518, 0.600] & 77.02 [72.30, 80.74] \\
Ours & 0.905 [0.886, 0.920] & 0.833 [0.806, 0.869] & 9.50 [8.51, 10.45] \\
\hline
\end{tabular*}
\end{center}
\begin{center}
\renewcommand{\arraystretch}{1.2}
\setlength{\tabcolsep}{3pt}
\begin{tabular*}{\textwidth}{@{\extracolsep{\fill}}lcccc@{}}
\hline
Model & $F_{\mathrm{On}}\uparrow$ & $F_{\mathrm{P37}}\uparrow$ & $F_{\mathrm{P13}}\uparrow$ & $F_{\mathrm{OpenMIC}}\uparrow$ \\
\hline
SpecDiff & 50.50 [42.64, 58.63] & 34.62 [26.44, 42.78] & 44.80 [38.26, 50.78] & 69.19 [53.08, 84.47] \\
CTD$^\dagger$ & 35.39 [26.62, 43.60] & 2.89 [0.57, 6.28] & 22.56 [13.38, 32.71] & 6.67 [0.00, 16.67] \\
Ours & 54.24 [48.28, 60.37] & 37.85 [26.85, 49.78] & 47.90 [40.66, 55.63] & 53.00 [32.00, 74.12] \\
\hline
\end{tabular*}
\end{center}

\clearpage

\subsection{GuitarSet}
\begin{center}
\renewcommand{\arraystretch}{1.2}
\setlength{\tabcolsep}{3pt}
\begin{tabular*}{\textwidth}{@{\extracolsep{\fill}}lccc@{}}
\hline
Model & MuQ$_{\mathrm{eval}}\uparrow$ & PQ$\uparrow$ & FAD$_{\mathrm{MERT}}\downarrow$ \\
\hline
SpecDiff & 4.06 [3.95, 4.17] & 7.89 [7.75, 8.02] & 5.92 \\
CTD & 3.56 [3.48, 3.64] & 7.95 [7.80, 8.08] & 8.67 \\
TokenSynth & 3.70 [3.49, 3.91] & 7.33 [6.98, 7.67] & 14.77 \\
Ours & 4.14 [4.01, 4.27] & 8.40 [8.33, 8.46] & 4.05 \\
\hline
\end{tabular*}
\end{center}
\begin{center}
\renewcommand{\arraystretch}{1.2}
\setlength{\tabcolsep}{3pt}
\begin{tabular*}{\textwidth}{@{\extracolsep{\fill}}lccc@{}}
\hline
Model & MuQ$_{\mathrm{cos}}\uparrow$ & CLAP$_{\mathrm{cos}}\uparrow$ & MSS$\downarrow$ \\
\hline
SpecDiff & 0.892 [0.881, 0.903] & 0.710 [0.686, 0.733] & 17.76 [14.43, 21.21] \\
CTD & 0.756 [0.723, 0.788] & 0.574 [0.503, 0.646] & 48.41 [42.98, 54.31] \\
TokenSynth & 0.686 [0.630, 0.734] & 0.367 [0.279, 0.457] & 110.17 [95.44, 125.50] \\
Ours & 0.921 [0.913, 0.929] & 0.908 [0.890, 0.923] & 14.29 [11.28, 17.46] \\
\hline
\end{tabular*}
\end{center}
\begin{center}
\renewcommand{\arraystretch}{1.2}
\setlength{\tabcolsep}{3pt}
\begin{tabular*}{\textwidth}{@{\extracolsep{\fill}}lcc@{}}
\hline
Model & $F_{\mathrm{On}}\uparrow$ & $F_{\mathrm{OpenMIC}}\uparrow$ \\
\hline
SpecDiff & 83.67 [75.22, 90.60] & 80.83 [71.67, 89.17] \\
CTD & 47.37 [37.99, 57.23] & 20.00 [5.00, 40.00] \\
TokenSynth & 34.36 [25.30, 44.18] & 11.67 [0.00, 25.00] \\
Ours & 88.01 [83.50, 92.08] & 85.00 [70.00, 100.00] \\
\hline
\end{tabular*}
\end{center}

\subsection{MAESTRO}
\begin{center}
\renewcommand{\arraystretch}{1.2}
\setlength{\tabcolsep}{3pt}
\begin{tabular*}{\textwidth}{@{\extracolsep{\fill}}lccc@{}}
\hline
Model & MuQ$_{\mathrm{eval}}\uparrow$ & PQ$\uparrow$ & FAD$_{\mathrm{MERT}}\downarrow$ \\
\hline
SpecDiff & 4.47 [4.36, 4.56] & 6.58 [6.44, 6.72] & 4.46 \\
CTD & 3.75 [3.60, 3.90] & 6.89 [6.65, 7.12] & 6.82 \\
TokenSynth & 3.95 [3.82, 4.10] & 6.99 [6.78, 7.19] & 9.00 \\
MIDI-VALLE & 4.37 [4.22, 4.52] & 7.64 [7.56, 7.72] & 3.67 \\
U-MusT & 4.35 [4.21, 4.49] & 7.24 [7.03, 7.43] & 2.59 \\
Ours & 4.61 [4.46, 4.75] & 7.50 [7.38, 7.60] & 2.89 \\
\hline
\end{tabular*}
\end{center}
\begin{center}
\renewcommand{\arraystretch}{1.2}
\setlength{\tabcolsep}{3pt}
\begin{tabular*}{\textwidth}{@{\extracolsep{\fill}}lccc@{}}
\hline
Model & MuQ$_{\mathrm{cos}}\uparrow$ & CLAP$_{\mathrm{cos}}\uparrow$ & MSS$\downarrow$ \\
\hline
SpecDiff & 0.925 [0.914, 0.936] & 0.705 [0.670, 0.740] & 31.24 [27.21, 35.15] \\
CTD & 0.741 [0.715, 0.766] & 0.662 [0.623, 0.699] & 49.22 [46.47, 52.13] \\
TokenSynth & 0.785 [0.758, 0.813] & 0.473 [0.430, 0.515] & 63.15 [57.00, 70.76] \\
MIDI-VALLE & 0.932 [0.920, 0.943] & 0.863 [0.826, 0.891] & 23.56 [20.32, 26.67] \\
U-MusT & 0.915 [0.899, 0.930] & 0.858 [0.825, 0.888] & 30.58 [26.04, 35.22] \\
Ours & 0.964 [0.958, 0.970] & 0.863 [0.831, 0.888] & 20.23 [17.51, 22.94] \\
\hline
\end{tabular*}
\end{center}
\begin{center}
\renewcommand{\arraystretch}{1.2}
\setlength{\tabcolsep}{3pt}
\begin{tabular*}{\textwidth}{@{\extracolsep{\fill}}lcc@{}}
\hline
Model & $F_{\mathrm{On}}\uparrow$ & $F_{\mathrm{OpenMIC}}\uparrow$ \\
\hline
SpecDiff & 66.63 [59.01, 74.06] & 100.00 [100.00, 100.00] \\
CTD & 16.67 [13.78, 19.87] & 96.67 [91.67, 100.00] \\
TokenSynth & 22.90 [19.07, 26.84] & 60.00 [48.29, 71.67] \\
MIDI-VALLE & 58.27 [51.13, 65.43] & 100.00 [100.00, 100.00] \\
U-MusT & 36.73 [30.12, 43.46] & 100.00 [100.00, 100.00] \\
Ours & 76.49 [70.54, 81.74] & 100.00 [100.00, 100.00] \\
\hline
\end{tabular*}
\end{center}
\clearpage
\section{Editing Results with Confidence Intervals (Table 2)}
\label{app:editing_ci}
The reporting convention in Appendix~\ref{app:synthesis_ci} also applies here. Each original song is one bootstrap cluster, keeping its related edit cases, target durations, and dry/wet renders together. Within each group, all models are evaluated on the same pairs. Per-example scores are averaged over the relevant pairs in each group, as specified below. Ours and Ours + FlowEdit use the same Base SQ codec. The dagger retains its meaning from Table~\ref{tab:edit_trial}.

\noindent\textbf{Slakh.} The editing set contains 668 original--revised pairs. Each 20.48-second crop contains an editing region spanning 30\%, 50\%, or 70\% of its duration. Track insertion, reported as ``Add new instrument'', comprises 100 drum-inclusive pairs from 100 songs. ``Note ins/del'' pools note insertion, note deletion, and track deletion: 93/91/91 pairs without drums and 97/98/98 with drums, respectively. These give 275 pairs from 127 songs and 293 pairs from 129 songs. In the two note insertion/deletion groups, $R_{\mathrm{add}}$ uses the note-insertion cases, while $R_{\mathrm{del}}$ uses the note- and track-deletion cases (182 and 196 pairs). $F_{\mathrm{keep}}$ uses cases with unchanged notes (273 and 293 pairs). A dash indicates that no deletions were requested.

\noindent\textbf{POP909.} Modulator supplies revised MIDI for 5-, 7-, and 10-second editing regions within 20.48-second crops. Each of 147 songs contributes one original--revised pair per duration and rendering condition (dry or matched reverberation), giving 882 pairs. Per-example scores are averaged over pairs from all six conditions. $F_{\mathrm{keep}}$ uses the 874 pairs with unchanged notes.
\subsection{Slakh (+drums): add new instrument}
\begin{center}
\renewcommand{\arraystretch}{1.2}
\setlength{\tabcolsep}{3pt}
\begin{tabular*}{\textwidth}{@{\extracolsep{\fill}}lccc@{}}
\hline
Model & MuQ$_{\mathrm{eval}}\uparrow$ & PQ$\uparrow$ & FAD$_{\mathrm{MERT}}\downarrow$ \\
\hline
SpecDiff & 4.29 [4.22, 4.36] & 7.71 [7.63, 7.79] & 4.87 \\
U-MusT & 3.72 [3.64, 3.79] & 7.98 [7.92, 8.04] & 3.45 \\
Ours & 4.44 [4.36, 4.52] & 8.13 [8.08, 8.18] & 1.68 \\
Ours + FlowEdit & 4.34 [4.26, 4.42] & 8.01 [7.94, 8.07] & 1.43 \\
\hline
\end{tabular*}
\end{center}
\begin{center}
\renewcommand{\arraystretch}{1.2}
\setlength{\tabcolsep}{3pt}
\begin{tabular*}{\textwidth}{@{\extracolsep{\fill}}lccc@{}}
\hline
Model & MuQ$_{\mathrm{cos}}\uparrow$ & CLAP$_{\mathrm{cos}}\uparrow$ & MSS$\downarrow$ \\
\hline
SpecDiff & 0.787 [0.773, 0.801] & 0.631 [0.610, 0.650] & 45.04 [42.47, 48.03] \\
U-MusT & 0.699 [0.680, 0.717] & 0.757 [0.739, 0.775] & 41.39 [38.94, 43.96] \\
Ours & 0.874 [0.863, 0.884] & 0.845 [0.829, 0.860] & 31.54 [30.09, 32.98] \\
Ours + FlowEdit & 0.882 [0.872, 0.892] & 0.833 [0.815, 0.850] & 24.40 [22.95, 25.92] \\
\hline
\end{tabular*}
\end{center}
\begin{center}
\renewcommand{\arraystretch}{1.2}
\setlength{\tabcolsep}{3pt}
\begin{tabular*}{\textwidth}{@{\extracolsep{\fill}}lcccc@{}}
\hline
Model & $R_{\mathrm{add}}\uparrow$ & $R_{\mathrm{del}}\downarrow$ & $F_{\mathrm{keep}}\uparrow$ & $F_{\mathrm{OpenMIC}}\uparrow$ \\
\hline
SpecDiff & 43.68 [38.22, 49.15] & \textemdash & 32.54 [29.15, 35.92] & 12.86 [9.95, 15.86] \\
U-MusT & 21.32 [18.07, 24.78] & \textemdash & 8.34 [6.75, 9.96] & 11.66 [8.63, 14.81] \\
Ours & 49.13 [43.55, 54.65] & \textemdash & 31.09 [27.87, 34.33] & 14.73 [11.45, 18.10] \\
Ours + FlowEdit & 42.14 [36.64, 47.79] & \textemdash & 32.68 [29.12, 36.24] & 14.85 [11.63, 18.00] \\
\hline
\end{tabular*}
\end{center}

\subsection{Slakh (without drums): note insertion/deletion}
\begin{center}
\renewcommand{\arraystretch}{1.2}
\setlength{\tabcolsep}{3pt}
\begin{tabular*}{\textwidth}{@{\extracolsep{\fill}}lccc@{}}
\hline
Model & MuQ$_{\mathrm{eval}}\uparrow$ & PQ$\uparrow$ & FAD$_{\mathrm{MERT}}\downarrow$ \\
\hline
CTD$^\dagger$ & 3.92 [3.85, 3.99] & 7.28 [7.19, 7.36] & 1.82 \\
Ours & 4.40 [4.33, 4.46] & 7.88 [7.83, 7.93] & 1.25 \\
Ours + FlowEdit & 4.32 [4.25, 4.39] & 7.76 [7.69, 7.83] & 1.06 \\
\hline
\end{tabular*}
\end{center}
\begin{center}
\renewcommand{\arraystretch}{1.2}
\setlength{\tabcolsep}{3pt}
\begin{tabular*}{\textwidth}{@{\extracolsep{\fill}}lccc@{}}
\hline
Model & MuQ$_{\mathrm{cos}}\uparrow$ & CLAP$_{\mathrm{cos}}\uparrow$ & MSS$\downarrow$ \\
\hline
CTD$^\dagger$ & 0.817 [0.807, 0.827] & 0.722 [0.709, 0.735] & 80.55 [76.79, 84.43] \\
Ours & 0.897 [0.890, 0.903] & 0.840 [0.830, 0.851] & 21.37 [20.30, 22.43] \\
Ours + FlowEdit & 0.911 [0.904, 0.918] & 0.861 [0.849, 0.872] & 11.15 [10.52, 11.79] \\
\hline
\end{tabular*}
\end{center}
\begin{center}
\renewcommand{\arraystretch}{1.2}
\setlength{\tabcolsep}{3pt}
\begin{tabular*}{\textwidth}{@{\extracolsep{\fill}}lcccc@{}}
\hline
Model & $R_{\mathrm{add}}\uparrow$ & $R_{\mathrm{del}}\downarrow$ & $F_{\mathrm{keep}}\uparrow$ & $F_{\mathrm{OpenMIC}}\uparrow$ \\
\hline
CTD$^\dagger$ & 43.20 [37.98, 48.28] & 3.82 [2.80, 4.98] & 28.91 [26.64, 31.23] & 17.33 [14.21, 20.51] \\
Ours & 42.51 [36.31, 48.78] & 6.08 [4.32, 8.05] & 33.24 [30.65, 35.84] & 26.56 [23.29, 29.78] \\
Ours + FlowEdit & 43.49 [37.19, 49.84] & 8.01 [6.16, 9.96] & 35.75 [33.16, 38.46] & 24.56 [21.22, 27.95] \\
\hline
\end{tabular*}
\end{center}

\clearpage

\subsection{Slakh (+drums): note insertion/deletion}
\begin{center}
\renewcommand{\arraystretch}{1.2}
\setlength{\tabcolsep}{3pt}
\begin{tabular*}{\textwidth}{@{\extracolsep{\fill}}lccc@{}}
\hline
Model & MuQ$_{\mathrm{eval}}\uparrow$ & PQ$\uparrow$ & FAD$_{\mathrm{MERT}}\downarrow$ \\
\hline
SpecDiff & 4.25 [4.19, 4.31] & 7.72 [7.65, 7.78] & 4.15 \\
U-MusT & 3.72 [3.67, 3.78] & 7.95 [7.89, 8.00] & 2.65 \\
Ours & 4.44 [4.39, 4.50] & 8.03 [7.98, 8.08] & 1.03 \\
Ours + FlowEdit & 4.36 [4.31, 4.42] & 7.93 [7.87, 7.99] & 0.83 \\
\hline
\end{tabular*}
\end{center}
\begin{center}
\renewcommand{\arraystretch}{1.2}
\setlength{\tabcolsep}{3pt}
\begin{tabular*}{\textwidth}{@{\extracolsep{\fill}}lccc@{}}
\hline
Model & MuQ$_{\mathrm{cos}}\uparrow$ & CLAP$_{\mathrm{cos}}\uparrow$ & MSS$\downarrow$ \\
\hline
SpecDiff & 0.797 [0.787, 0.806] & 0.619 [0.605, 0.632] & 40.33 [38.86, 41.89] \\
U-MusT & 0.699 [0.685, 0.712] & 0.744 [0.730, 0.756] & 37.81 [36.31, 39.38] \\
Ours & 0.894 [0.885, 0.902] & 0.870 [0.858, 0.881] & 23.26 [22.32, 24.22] \\
Ours + FlowEdit & 0.913 [0.905, 0.920] & 0.882 [0.872, 0.891] & 12.65 [12.05, 13.31] \\
\hline
\end{tabular*}
\end{center}
\begin{center}
\renewcommand{\arraystretch}{1.2}
\setlength{\tabcolsep}{3pt}
\begin{tabular*}{\textwidth}{@{\extracolsep{\fill}}lcccc@{}}
\hline
Model & $R_{\mathrm{add}}\uparrow$ & $R_{\mathrm{del}}\downarrow$ & $F_{\mathrm{keep}}\uparrow$ & $F_{\mathrm{OpenMIC}}\uparrow$ \\
\hline
SpecDiff & 52.64 [46.67, 58.47] & 2.62 [1.67, 3.80] & 39.19 [36.44, 41.91] & 15.48 [13.07, 17.98] \\
U-MusT & 20.50 [16.53, 24.67] & 3.72 [2.83, 4.72] & 9.56 [8.45, 10.82] & 13.28 [11.25, 15.43] \\
Ours & 56.44 [50.64, 62.13] & 4.56 [3.21, 6.24] & 41.90 [39.34, 44.52] & 16.92 [14.61, 19.28] \\
Ours + FlowEdit & 50.54 [44.85, 56.27] & 7.78 [5.99, 9.71] & 41.55 [38.90, 44.22] & 16.21 [13.71, 18.78] \\
\hline
\end{tabular*}
\end{center}

\subsection{POP909: AI edits}
\begin{center}
\renewcommand{\arraystretch}{1.2}
\setlength{\tabcolsep}{3pt}
\begin{tabular*}{\textwidth}{@{\extracolsep{\fill}}lccc@{}}
\hline
Model & MuQ$_{\mathrm{eval}}\uparrow$ & PQ$\uparrow$ & FAD$_{\mathrm{MERT}}\downarrow$ \\
\hline
SpecDiff & 4.35 [4.31, 4.39] & 7.47 [7.42, 7.51] & 5.46 \\
CTD & 3.66 [3.62, 3.70] & 7.42 [7.38, 7.46] & 7.99 \\
TokenSynth & 3.79 [3.75, 3.83] & 7.41 [7.38, 7.45] & 9.21 \\
MIDI-VALLE & 4.13 [4.09, 4.16] & 7.95 [7.94, 7.97] & 5.19 \\
U-MusT & 4.26 [4.22, 4.29] & 7.78 [7.76, 7.80] & 6.50 \\
Ours & 4.42 [4.38, 4.45] & 7.99 [7.97, 8.01] & 3.33 \\
Ours + FlowEdit & 4.36 [4.33, 4.40] & 8.03 [8.01, 8.05] & 2.31 \\
\hline
\end{tabular*}
\end{center}
\begin{center}
\renewcommand{\arraystretch}{1.2}
\setlength{\tabcolsep}{3pt}
\begin{tabular*}{\textwidth}{@{\extracolsep{\fill}}lccc@{}}
\hline
Model & MuQ$_{\mathrm{cos}}\uparrow$ & CLAP$_{\mathrm{cos}}\uparrow$ & MSS$\downarrow$ \\
\hline
SpecDiff & 0.873 [0.869, 0.876] & 0.762 [0.755, 0.768] & 30.92 [29.59, 32.31] \\
CTD & 0.716 [0.708, 0.723] & 0.652 [0.642, 0.662] & 45.11 [43.91, 46.36] \\
TokenSynth & 0.741 [0.732, 0.749] & 0.625 [0.617, 0.634] & 87.71 [85.24, 90.29] \\
MIDI-VALLE & 0.865 [0.861, 0.869] & 0.729 [0.720, 0.737] & 23.93 [22.48, 25.35] \\
U-MusT & 0.842 [0.838, 0.846] & 0.810 [0.802, 0.818] & 29.00 [27.84, 30.27] \\
Ours & 0.920 [0.918, 0.922] & 0.766 [0.761, 0.772] & 16.82 [15.99, 17.65] \\
Ours + FlowEdit & 0.915 [0.912, 0.917] & 0.777 [0.771, 0.783] & 15.64 [14.78, 16.52] \\
\hline
\end{tabular*}
\end{center}
\begin{center}
\renewcommand{\arraystretch}{1.2}
\setlength{\tabcolsep}{3pt}
\begin{tabular*}{\textwidth}{@{\extracolsep{\fill}}lcccc@{}}
\hline
Model & $R_{\mathrm{add}}\uparrow$ & $R_{\mathrm{del}}\downarrow$ & $F_{\mathrm{keep}}\uparrow$ & $F_{\mathrm{OpenMIC}}\uparrow$ \\
\hline
SpecDiff & 83.03 [81.06, 84.78] & 2.60 [2.25, 2.96] & 12.85 [11.12, 14.67] & 98.85 [97.51, 99.69] \\
CTD & 30.16 [28.43, 31.83] & 6.74 [6.06, 7.43] & 3.23 [2.73, 3.80] & 97.03 [96.11, 97.88] \\
TokenSynth & 27.24 [25.93, 28.54] & 5.27 [4.77, 5.79] & 7.66 [6.95, 8.47] & 80.50 [78.63, 82.39] \\
MIDI-VALLE & 72.74 [71.05, 74.38] & 4.86 [4.36, 5.40] & 10.41 [9.12, 11.77] & 99.74 [99.24, 100.00] \\
U-MusT & 42.56 [40.88, 44.28] & 5.39 [4.88, 5.92] & 3.77 [3.21, 4.33] & 99.96 [99.89, 100.00] \\
Ours & 89.86 [88.93, 90.71] & 3.40 [3.07, 3.74] & 13.84 [11.96, 15.72] & 99.81 [99.58, 100.00] \\
Ours + FlowEdit & 86.81 [85.81, 87.77] & 5.06 [4.58, 5.55] & 12.46 [10.71, 14.30] & 99.89 [99.70, 100.00] \\
\hline
\end{tabular*}
\end{center}
\clearpage
\section{Note-Adherence Error Analysis with Confidence Intervals (Table 3)}
\label{app:codec_ci}
Both panels use Slakh (+drums), with 100 examples from 100 songs per panel. All scores are percentages evaluated with YourMT3+ at a 50-ms onset tolerance. Ground-truth audio, continuous reconstruction without SQ, and Base SQ reconstruction use the same target MIDI as the generated output. Panel (b) evaluates the revised ground-truth audio and its reconstructions against the requested edit. Perfect AMT is a theoretical 100\% control, not an empirical estimate, and therefore has no CI.

\subsection{Synthesis}
\begin{center}
\renewcommand{\arraystretch}{1.2}
\setlength{\tabcolsep}{3pt}
\begin{tabular*}{\textwidth}{@{\extracolsep{\fill}}lcc@{}}
\hline
Evaluated audio / transcription & $F_{\mathrm{P37}}\uparrow$ & $F_{\mathrm{On}}\uparrow$ \\
\hline
Ground Truth (GT) w/ Perfect AMT & 100.00 & 100.00 \\
GT w/ AMT & 78.40 [76.58, 80.16] & 82.79 [81.32, 84.21] \\
Reconstruction (no SQ) w/ AMT & 44.43 [41.90, 46.93] & 55.94 [53.59, 58.27] \\
Reconstruction (SQ) w/ AMT & 43.98 [41.36, 46.57] & 55.79 [53.48, 58.01] \\
Generated (ours) & 37.78 [35.17, 40.40] & 54.06 [51.59, 56.48] \\
\hline
\end{tabular*}
\end{center}
\subsection{Add new instrument}
\begin{center}
\renewcommand{\arraystretch}{1.2}
\setlength{\tabcolsep}{3pt}
\begin{tabular*}{\textwidth}{@{\extracolsep{\fill}}lcc@{}}
\hline
Evaluated audio / transcription & $R_{\mathrm{add}}\uparrow$ & $F_{\mathrm{keep}}\uparrow$ \\
\hline
Ground Truth (GT) w/ Perfect AMT & 100.00 & 100.00 \\
GT w/ AMT & 71.40 [66.79, 75.86] & 72.80 [69.92, 75.55] \\
Reconstruction (no SQ) w/ AMT & 55.99 [50.95, 60.89] & 39.01 [35.33, 42.71] \\
Reconstruction (SQ) w/ AMT & 54.54 [49.65, 59.44] & 37.50 [33.88, 41.21] \\
Generated (ours) & 49.13 [43.55, 54.65] & 31.09 [27.87, 34.33] \\
\hline
\end{tabular*}
\end{center}
\section{Inference-Time Ablations with Confidence Intervals (Table 4)}
\label{app:ablation_ci}
The same trained checkpoint is evaluated on 56 target intervals, eight from one recording each in Slakh, MAESTRO, PianoVAM, MusicNetEM, URMP, KRAISLER, and FiloBass (dataset references in Appendix~\ref{app:datasets}). The bootstrap resamples these seven recordings, keeping their intervals together. The intervals describe uncertainty for this small, mixed-dataset collection and should not be interpreted as dataset-specific estimates.

Base uses contextual MIDI (CM), the clean contextual audio condition $\mathbf A$ (CA), and CFG~2. The 32-step ablations remove CM, zero only $\mathbf A$, or disable CFG (CFG~1), without retraining. The observed frames in the noisy input remain available in the $-$CA condition. FAD uses pooled features from all 56 intervals and is reported without a CI.
\begin{center}
\renewcommand{\arraystretch}{1.2}
\setlength{\tabcolsep}{3pt}
\begin{tabular*}{\textwidth}{@{\extracolsep{\fill}}lccc@{}}
\hline
Setting & MuQ$_{\mathrm{eval}}\uparrow$ & FAD$_{\mathrm{MERT}}\downarrow$ & $F_{\mathrm{On}}\uparrow$ \\
\hline
Base, 4 steps & 4.48 [4.27, 4.68] & 3.65 & 74.60 [59.55, 86.92] \\
Base, 8 steps & 4.46 [4.26, 4.65] & 3.29 & 76.14 [62.07, 87.79] \\
Base, 16 steps & 4.43 [4.24, 4.61] & 3.09 & 75.90 [61.03, 87.61] \\
Base, 32 steps & 4.41 [4.22, 4.60] & 2.99 & 73.34 [58.36, 85.86] \\
Base, 64 steps & 4.40 [4.22, 4.57] & 2.95 & 73.03 [58.28, 85.92] \\
$-$CM, 32 steps & 4.44 [4.26, 4.62] & 3.36 & 76.01 [62.27, 87.04] \\
$-$CA, 32 steps & 3.48 [3.00, 3.90] & 9.95 & 45.66 [29.92, 65.58] \\
$-$CFG, 32 steps & 4.29 [4.11, 4.44] & 2.98 & 65.67 [51.17, 79.98] \\
\hline
\end{tabular*}
\end{center}

\clearpage
\makeatletter
\renewenvironment{thebibliography}[1]{%
  \section*{\centering\fontsize{9}{11}\selectfont APPENDIX REFERENCES}
  \list{[A\arabic{enumi}]}{\settowidth\labelwidth{[A#1]}%
    \leftmargin\labelwidth\advance\leftmargin\labelsep\usecounter{enumi}}%
  \def\newblock{\hskip .11em plus .33em minus .07em}%
  \sloppy\clubpenalty4000\widowpenalty4000\sfcode`\.=1000\relax
}{\endlist}
\makeatother
\putbib[refs]
\end{bibunit}

\end{document}